\documentclass[11pt]{article}
\pdfoutput=1
\usepackage{amsmath,amssymb}
\usepackage{graphicx,color}
\usepackage{hyperref}

\usepackage{url}

\allowdisplaybreaks[2]

\def\cW{\mathcal{W}}

\def\mint{\int_{-\infty}^\infty\!\cdots\!\int_{-\infty}^\infty}

\newcommand{\be}{\begin{equation}}
\newcommand{\ee}{\end{equation}}
\newcommand{\ba}{\begin{aligned}}
\newcommand{\ea}{\end{aligned}}

\def\vev#1{\langle #1 \rangle}

\def\bra#1{\left\langle #1 \right|}
\def\ket#1{\left| #1 \right\rangle}

\def\({\left(}
\def\){\right)}

\DeclareMathOperator{\Tr}{Tr}

\newcommand{\ri}{{\rm i}}

\def\th{\theta}

\newcommand{\hf}{\frac{1}{2}}

\def\del{\partial}

\def\bra{\langle}
\def\ket{\rangle}

\def\la{\lambda}

\def\al{\alpha}

\def\rt#1{\sqrt{#1}}
\def\sitarel#1#2{\mathrel{\mathop{\kern0pt #1}\limits_{#2}}}

\makeatletter

\@addtoreset{equation}{section}
\makeatother

\title{\bf $\alpha'$-expansion of Anti-Symmetric Wilson Loops\\ in $\mathcal{N}=4$ SYM from Fermi Gas}

\author{Masaatsu Horikoshi and Kazumi Okuyama}
\date{Department of Physics, 
Shinshu University,\\
 Matsumoto 390-8621, Japan}

\begin{document}

\maketitle
\renewcommand{\thefootnote}{\arabic{footnote}}
\setcounter{footnote}{0}
\setcounter{section}{0}

\begin{abstract}
We study the large 't Hooft coupling expansion
of 1/2 BPS Wilson loops in the anti-symmetric representation
in $\mathcal{N}=4$ super Yang-Mills (SYM)
theory
at the leading order in the $1/N$ expansion.
Via AdS/CFT correspondence,
this expansion corresponds to the $\alpha'$ expansion
in bulk type IIB string theory. 
We show that this expansion
can be systematically computed by using the 
low temperature expansion of 
Fermi distribution function, known as the Sommerfeld expansion
in statistical mechanics. 
We check numerically that our expansion 
agrees with the
exact result of anti-symmetric Wilson loops 
recently found by Fiol and Torrents. 
\end{abstract}

\newpage
\section{Introduction}
1/2 BPS circular Wilson loops in
4d $\mathcal{N}=4$ super Yang-Mills (SYM) theory 
are interesting observables which can be computed exactly
by a Gaussian matrix model \cite{Erickson:2000af,Drukker:2000rr,Pestun:2007rz}.
Via AdS/CFT correspondence,
1/2 BPS Wilson loops in the fundamental
representation correspond to the fundamental string
in type IIB string theory on $AdS_5\times S^5$ \cite{Maldacena:1998im,Rey:1998ik}.
When the rank of the representations becomes large,
the corresponding dual objects in the bulk are not fundamental strings but D-branes, and such Wilson loops are sometimes called ``Giant Wilson loops''.
In particular,
1/2 BPS Wilson loops in the rank $k$
symmetric and anti-symmetric representations
correspond to D3-branes and D5-branes, respectively,
with $k$ unit of electric flux on their worldvolumes \cite{Drukker:2005kx,Yamaguchi:2006tq,Hartnoll:2006is,Okuyama:2006jc}.
The leading term in the 't Hooft expansion
of $\mathcal{N}=4$ SYM side
is successfully matched
with the DBI action of D-branes in the bulk side.
For more general representations, a dictionary
between Wilson loops in higher rank representations
and bulk D-brane picture was proposed in \cite{Gomis:2006sb,Gomis:2006im}\footnote{
Bubbling geometries including the effect of back-reaction due to
the insertion of Wilson loops were studied in
\cite{Yamaguchi:2006te,Lunin:2006xr}.}.

We are interested in the subleading corrections
in this correspondence.
Recently, there are some progress in the 
computation of one-loop correction in the $1/N$ expansion of Giant Wilson loops
\cite{Faraggi:2014tna,Buchbinder:2014nia,Faraggi:2011ge,Faraggi:2011bb}.
Here we will focus on the subleading corrections
in the large $\la$ expansion (or $1/\la$ expansion)
with $\la$ being the 't Hooft coupling $\la=g_{\text{YM}}^2N$,
and we will restrict ourselves to the leading order
   in the $1/N$ expansion.
   From the holographic dictionary $R_{AdS}^2/\al'=\rt{\la}$,
   the large $\la$ expansion on the SYM side
   corresponds
   to the $\al'$ expansion in the bulk string theory side.

   In this paper, we consider the large $\la$ expansion
   of the 1/2 BPS Wilson loops in the anti-symmetric representation.
   Using the fact that the generating function of
   anti-symmetric representations can be written
   as a system of fermions, one can systematically compute the
   subleading corrections in the large $\la$ expansion
   by a low temperature expansion of the Fermi distribution
   function, known as the Sommerfeld expansion.
   Here the role of temperature is played by $1/\rt{\la}$.
   We have checked numerically that the  subleading corrections
   agree with the exact expression of the anti-symmetric
   Wilson loops recently found in \cite{Fiol:2013hna}.

   The rest of the paper is organized as follows.
   In section \ref{sec:fermi},
   we find a systematic large $\la$
   expansion of anti-symmetric
   Wilson loops using the Sommerfeld expansion of Fermi distribution function.
   Our main result is \eqref{eq:W-expand}.
   In section \ref{sec:compare}, we compare
   our result \eqref{eq:W-expand} with
   the exact expression in \cite{Fiol:2013hna},
   and find a nice agreement.
   We conclude in section \ref{sec:discuss}
   and discuss some future directions.
\section{Large $\la$ expansion of Wilson loops in the anti-symmetric representation}\label{sec:fermi}
We consider the vacuum expectation value (VEV)
of 1/2 BPS circular Wilson loops in $\mathcal{N}=4$ SYM
with gauge group $U(N)$.
After applying the supersymmetric localization
 \cite{Pestun:2007rz},
 the Wilson loop VEV is reduced to a Gaussian matrix model
\begin{align}
 \Biggl\bra \Tr_R P\exp\left[\oint ds
 \Bigl(\ri A_\mu\dot{x}^\mu+\Phi_I\th^I
 |\dot{x}|\Bigr)\right]\Biggr\ket=\int dM e^{-\frac{1}{2\pi g_s}\Tr M^2}\Tr_R (e^M).
\label{eq:mat-int}
\end{align}
Here $x^\mu(s)$ parametrizes a great circle of $S^4$ on which
$\mathcal{N}=4$ SYM lives, and
$\Phi_I~(I=1,\cdots,6)$ denote the adjoint scalar fields in $\mathcal{N}=4$ SYM
and $\th^I\in S^5$ is a constant unit vector. 
$g_s$ in \eqref{eq:mat-int} denotes the string coupling which is related to the
Yang-Mills gauge coupling $g_\text{YM}$ by
\begin{align}
 g_s=\frac{g_\text{YM}^2}{4\pi}.
\end{align}
In this paper, we will focus on the
Wilson loop VEV in the $k^\text{th}$ anti-symmetric representation
$R=A_k$
\begin{align}
 W_{A_k}=\int dM e^{-\frac{1}{2\pi g_s}\Tr M^2}\Tr_{A_k} (e^M).
\end{align}
It is convenient to define the VEV of $SU(N)$ part by removing the $U(1)$ contribution
\begin{align}
 \cW_{A_k}=W_{A_k}e^{-\frac{\pi kg_s}{2}}.
\end{align}
One can show that $\cW_{A_k}$ is symmetric under $k\to N-k$
\begin{align}
 \cW_{A_{N-k}}=\cW_{A_k}.
\end{align}

We are interested in the behavior of Wilson loop VEV $\cW_{A_k}$
in the limit
 \begin{align}
  N\to\infty\quad
  \text{with}\quad
  \lambda=g_\text{YM}^2N,~~\frac{k}{N}~~\text{fixed}.
  \label{eq:regime}
 \end{align}
 In the large $\la$ limit together with  \eqref{eq:regime},
the anti-symmetric Wilson loop $\cW_{A_k}$
 is holographically dual to a D5-brane in $AdS_5\times S^5$,
whose worldvolume
 has the form $AdS_2\times S^4$ \cite{Yamaguchi:2006tq}. From the computation of the DBI action of
 D5-brane,
 the leading behavior of $\cW_{A_k}$
 is found to be
  \begin{align}
\begin{aligned}
   \log \cW_{A_k}=\frac{2N\rt{\la}}{3\pi}\sin^3\th_k=
\frac{1}{g_s}\frac{(\rt{\la}\sin\th_k)^3}{6\pi^2},
\end{aligned}
  \end{align}
  where $\th_k$ is given by
\begin{align}
 \th_k-\sin\th_k\cos\th_k=\frac{\pi k}{N}. 
\end{align}
From the bulk D5-brane picture, the angle
$\th_k$ parametrizes
the position of $S^4$ part of the worldvolume
inside the $S^5$ of bulk geometry $AdS_5\times S^5$.

We are interested in the subleading corrections of $\cW_{A_k}$.
There are two expansion parameters $g_s$ and $1/\la$.
In \cite{Faraggi:2014tna}, it was reported that the
one-loop correction 
in the $g_s$ expansion has the form
\begin{align}
 \log \cW_{A_k}=\frac{1}{g_s}\frac{(\rt{\la}\sin\th_k)^3}{6\pi^2}
 +c\log\sin\th_k,
\label{eq:log-corr}
\end{align}
where $c$ is an order 1 constant.
In this paper, we will consider subleading corrections of the
$1/\la$ expansion while we focus on
the leading order in  the $g_s$-expansion.

As we will show below, the $1/\la$ expansion of
$\cW_{A_k}$ can be computed as
\begin{align}
\begin{aligned}
 \log \cW_{A_k}&=\frac{1}{g_s}
 \Biggl[\frac{(\rt{\la}\sin\th_k)^3}{6\pi^2}
 +\frac{\rt{\la}\sin\th_k}{12}
-\frac{\pi^2(19+5\cos2\th_k)}{\rt{\la}\sin^3\th_k}
\\
&\qquad-\frac{\pi ^4 (6788 \cos 2\th_k+35 \cos 4\th_k+8985)}{362880\la^{\frac{3}{2}}\sin^7\th_k}
+\cdots\Biggr].
\end{aligned}
 \label{eq:W-expand}
\end{align}
This is our main result.

Let us explain how we obtained \eqref{eq:W-expand}.
To study the anti-symmetric Wilson loops systematically,
it is convenient to introduce the generating function of
$\cW_{A_k}$ by summing over $k$ with fugacity $e^\mu$
\begin{align}
 \sum_{k=0}^N e^{k\mu} \cW_{A_k}=\vev{\det(1+e^\mu e^M)}_{mm}
\label{eq:gen-fun}
\end{align}
where $\vev{\mathcal{O}}_{mm}$ denotes the expectation value
in the Gaussian matrix model
\begin{align}
 \vev{\mathcal{O}}_{mm}=\int dM e^{-\frac{1}{2\pi g_s}\Tr M^2}\mathcal{O}.
\end{align}
Using the large $N$ factorization we find
\begin{align}
 \vev{\det(1+e^\mu e^M)}_{mm}\approx e^{\vev{\text{Tr}\log (1+e^\mu e^M)}_{mm}}
\label{eq:largeN-fac}
\end{align}
up to $1/N$ corrections, and the right hand side of \eqref{eq:largeN-fac}
in the planar limit becomes
\begin{align}
 \vev{\text{Tr}\log (1+e^\mu e^M)}_{mm}=N\int_{-\rt{\la}}^{\rt{\la}} dm\rho(m)\log(1+e^{\mu-m}),
\end{align}
where $\rho(m)$ is the Wigner semi-circle distribution
of Gaussian matrix model
\begin{align}
 \rho(m)=\frac{2}{\pi\la}\rt{\la-m^2}.
\end{align}
Then, as 
discussed in \cite{Hartnoll:2006is},  
the Wilson loop VEV in the $k^\text{th}$
anti-symmetric representation
is written as an integral over the chemical potential $\mu$
\begin{align}
 \cW_{A_k}=\int d\mu \exp\left[-k\mu +N\int_{-\rt{\la}}^{\rt{\la}} dm\rho(m)\log(1+e^{\mu-m})\right].
\label{eq:wmu1}
\end{align}
By rescaling $(m,\mu)\to(\rt{\la}m,\rt{\la}\mu)$, we can further rewrite
\eqref{eq:wmu1} as
\begin{align}
 \begin{aligned}
  \cW_{A_k}&=\int d\mu \exp\left[N\Biggl(
-\frac{k}{N}\rt{\la}\mu +
  J(\mu)\Biggr)\right],
 \end{aligned}
 \label{eq:mu-int}
\end{align}
where
 \begin{align}
  \begin{aligned}
   J(\mu)&=\frac{2}{\pi}\int_{-1}^1dm\rt{1-m^2}\log(1+e^{\rt{\la}(\mu-m)}).  
  \end{aligned}
 \end{align}
 In the regime of our interest \eqref{eq:regime}, 
 the $\mu$-integral in \eqref{eq:mu-int}
 can be evaluated by  the saddle point
  approximation since the exponent in \eqref{eq:mu-int}
is multiplied by the large number $N$.
  Thus we conclude that  $\log \cW_{A_k}$ is essentially given by
  the Legendre transform of $J(\mu)$
\begin{align}
 \log \cW_{A_k}=-k\rt{\la}\mu_*+NJ(\mu_*),
\label{eq:Legendre}
\end{align} 
where $\mu_*$ is determined by
the saddle point equation
  \begin{align}
   \del_\mu J(\mu)\Big|_{\mu=\mu_*}=\frac{k\rt{\la}}{N}.
\label{eq:saddle-pt}
  \end{align}
  Note that the fluctuation of $\mu$-integral around
  the saddle point gives rise to a subleading correction in $g_s$,
  as in the case of ABJM Fermi gas \cite{Marino:2011eh},
  and hence we can safely ignore such corrections for our purpose
to study the leading order behavior in the $g_s$ expansion\footnote{The overall constant of the integral
\eqref{eq:wmu1} and the factor coming from the change of variable
$\mu\to\rt{\la}\mu$ from \eqref{eq:wmu1} to \eqref{eq:mu-int} are also subleading
in the $g_s$ expansion, and we simply ignore them as well.}.

  Noticing that the Fermi distribution function naturally appears in
 the derivative of  $J(\mu)$
  \begin{align}
 \del_\mu J(\mu)=\frac{2\rt{\la}}{\pi}\int_{-1}^1dm \frac{\rt{1-m^2}}{1+e^{\rt{\la}(m-\mu)}},  
  \end{align}
 one can 
  easily compute the $1/\la$ expansion
  by the standard Sommerfeld expansion in statistical mechanics,
  where
  $1/\rt{\la}$ plays the role of temperature.

The large $\la$ expansion of Fermi distribution function reads
\begin{align}
\begin{aligned}
 \frac{1}{1+e^{\rt{\la}(m-\mu)}}&=\frac{\pi\del_\mu}{\rt{\la}\sin \frac{\pi\del_\mu}{\rt{\la}}}
\Theta(\mu-m)\\
&=\sum_{n=0}^\infty \frac{(-1)^nB_{2n}(1/2)}{(2n)!}\Biggl(\frac{4\pi^2\del_\mu^2}{\la}\Biggr)^{n}\Theta(\mu-m)
\\
&=\left(1+\frac{\pi^2\del_{\mu}^2}{6\la}
+\frac{7\pi^4\del_{\mu}^4}{360\la^2}+\cdots\right)\Theta(\mu-m) 
\end{aligned}
\end{align}
where $B_{2n}(1/2)$ is the value of  Bernoulli polynomial
$B_{2n}(z)$ at $z=1/2$, and 
$\Theta(\mu-m)$ is the step-function
\begin{align}
 \Theta(\mu-m)=\Biggl\{
\begin{aligned}
 1&\qquad(\mu>m),\\
0&\qquad(\mu<m).
\end{aligned}
\end{align}
Introducing the variable $\th$ by
  \begin{align}
 \mu=-\cos\th,
  \end{align}
   one can easily show that $\del_\mu J(\mu)$ is expanded as
 \begin{align}
\begin{aligned}
  \del_\mu J(\mu)
&=\frac{2\rt{\la}}{\pi}\Biggl[\hf(\th-\sin\th\cos\th)\\
&+\sum_{n=1}^\infty\frac{(-1)^nB_{2n}(1/2)}{(2n)!}\Biggl(\frac{4\pi^2}{\la}\Biggr)^{n}
\Biggl(\frac{1}{\sin\theta}\frac{\del}{\del\th}\Biggr)^{2n-1}\sin\th\Biggr],
\end{aligned}
 \end{align}
from which the expansion of $J(\mu)$ is found to be
\begin{align}
\begin{aligned}
 J(\mu)=&\int d\th \sin\th \del_\mu J(\mu)\\
=&
\frac{2\rt{\la}}{\pi}\Biggl[\frac{\sin^3\th}{3}-\hf(\th-\sin\th\cos\th)\cos\th\\
&+\sum_{n=1}^\infty\frac{(-1)^nB_{2n}(1/2)}{(2n)!}\Biggl(\frac{4\pi^2}{\la}\Biggr)^{n}
\Biggl(\frac{1}{\sin\theta}\frac{\del}{\del\th}\Biggr)^{2n-2}\sin\th\Biggr]. 
\end{aligned}
\end{align}
Finally, solving the saddle point equation \eqref{eq:saddle-pt}
order by order in
$1/\la$ expansion, and plugging the solution
$\mu_*$  into
\eqref{eq:Legendre}, we arrive at our main result
\eqref{eq:W-expand}. In this way, we can  compute the
$1/\la$ expansion of $\cW_{A_k}$ up to any desired order.

\section{Comparison with the exact result}\label{sec:compare}
Let us compare our result 
\eqref{eq:W-expand} with the exact result of anti-symmetric Wilson loops
at finite $N$ and $k$ found in \cite{Fiol:2013hna}.
\begin{figure}[tb]
\begin{center}
\begin{tabular}{cc}
\hspace{-4mm}
\includegraphics[width=7cm]{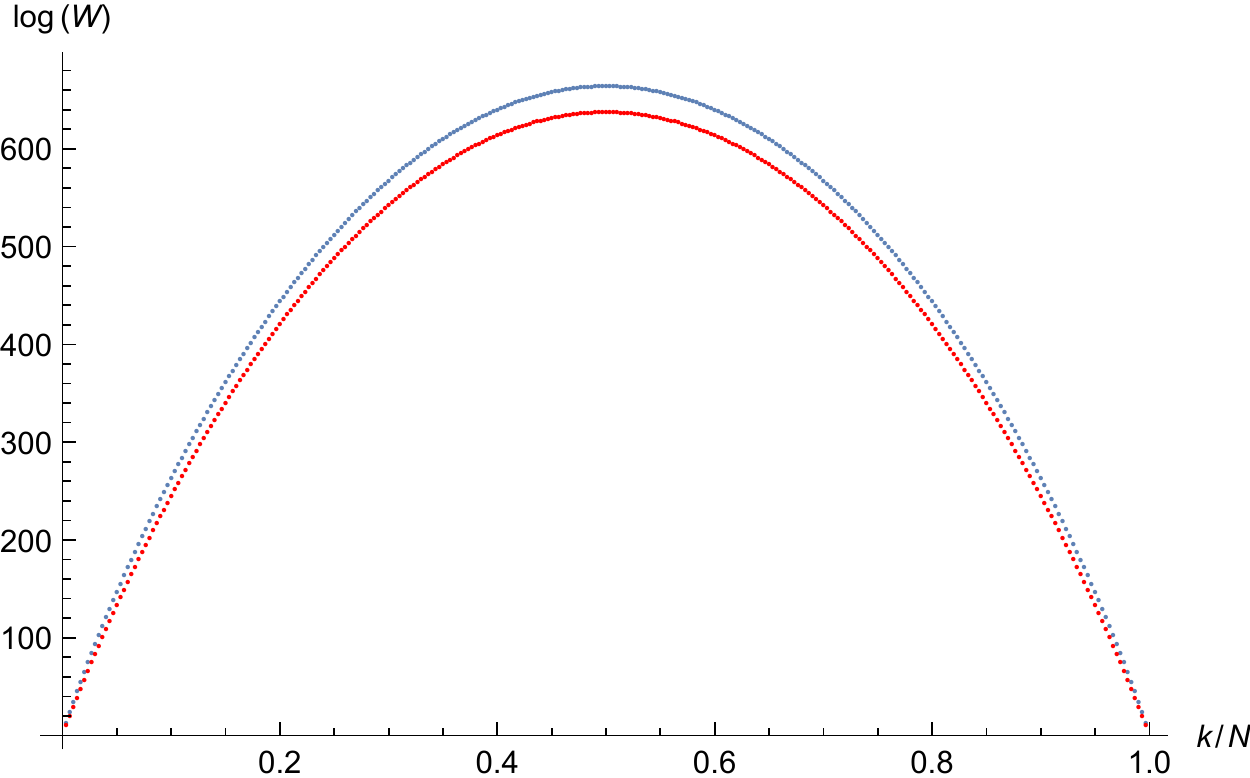}
\hspace{10mm}
&
\includegraphics[width=7cm]{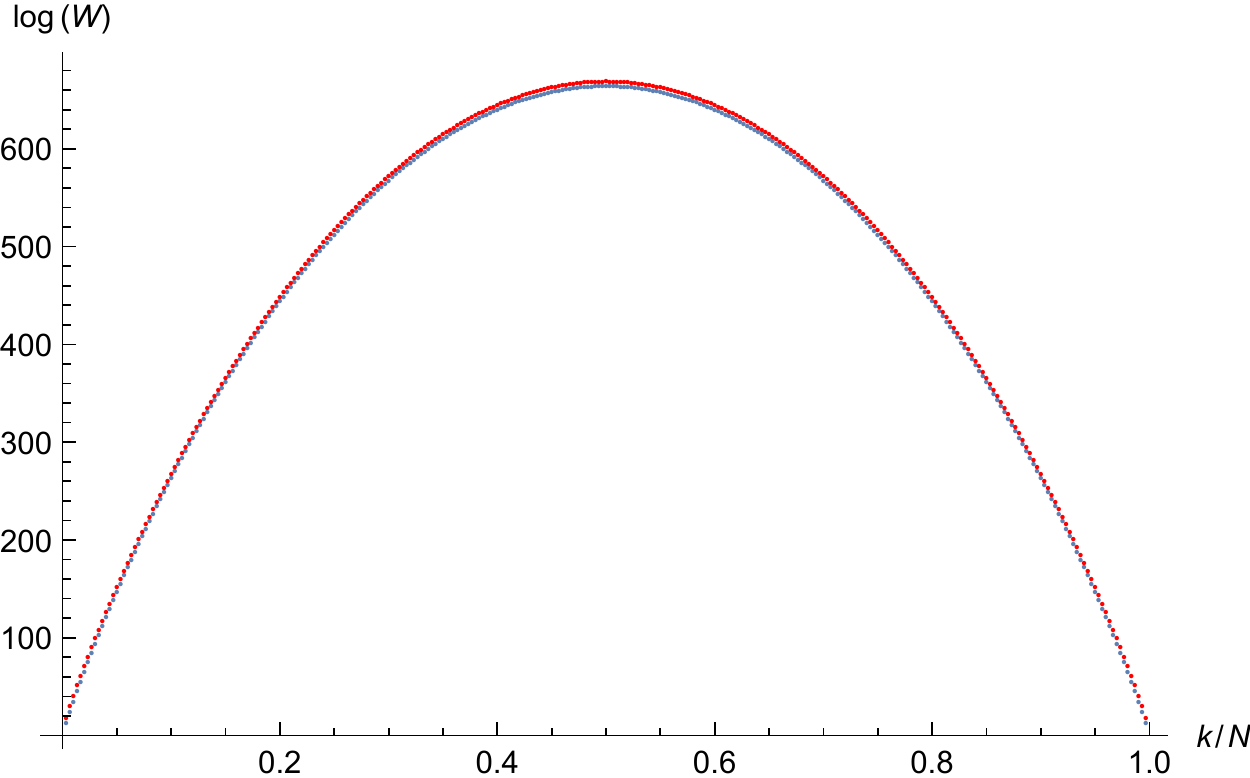}\\
(a) $\frac{1}{g_s}\frac{(\rt{\la}\sin\th_k)^3}{6\pi^2}$
\hspace{10mm}
&
(b) $\frac{1}{g_s}\left[\frac{(\rt{\la}\sin\th_k)^3}{6\pi^2}+\frac{\rt{\la}\sin\th_k}{12}\right]$ 
\end{tabular}
\end{center}
  \caption{This is  the plot of  $\log \cW_{A_k}$ as a function of $k/N$,
for $N=300$, $\la=100$. 
The blue dots are the exact values obtained from \eqref{eq:genA}, while
the red dots represent the
behavior of (a) leading term only (b) leading $+$ next-to-leading terms
in the expansion \eqref{eq:W-expand}.
One can clearly see that the inclusion of the next-to-leading
correction improves the matching.
}
  \label{fig:logW}
\end{figure}

It is found in \cite{Fiol:2013hna}
that the generating function of $\cW_{A_k}$ \eqref{eq:gen-fun}
is exactly written as
a characteristic polynomial of $N\times N$  matrix $A$ 
 \begin{align}  
  \sum_{k=0}^N {e^{k\mu}} \cW_{A_k}=\det(1+e^\mu A),
  \label{eq:genA}
 \end{align}
 where the matrix element $A_{i,j}$
is given by the generalized Laguerre polynomial
 \begin{align}
  A_{i,j}=L_{i-1}^{j-i}(-\pi g_s),\qquad(i,j=1,\cdots,N).
 \end{align}
From this expression  \eqref{eq:genA},
one can extract the exact value of Wilson loop VEV $\cW_{A_k}$ at arbitrary value of $N,k$ and $g_s$.

In Fig.\ref{fig:logW}, we show 
the plot of $\log \cW_{A_k}$  as a function of $k/N$
for $N=300$ and $\la=100$, corresponding to the
value of string coupling $g_s=\frac{\la}{4\pi N}=\frac{1}{12\pi}$.
The blue dots are the exact values obtained from \eqref{eq:genA}
while the red dots are the plot of our result \eqref{eq:W-expand}
for the leading term (Fig.\ref{fig:logW}(a)) and the leading $+$ next-to-leading terms (Fig.\ref{fig:logW}(b)).
One can clearly see that the 
leading $+$ next-to-leading  terms in Fig.\ref{fig:logW}(b)
exhibit a nice agreement with the exact result \eqref{eq:genA}.
Interestingly, the leading term alone is not enough to reproduce
the behavior of  the exact result \eqref{eq:genA},
and the next-to-leading correction has a rather large 
contribution for this choice of parameters $N=300,\la=100$.
Note that the leading and the next-to-leading terms in \eqref{eq:W-expand}
are of order $\la^{3/2}/g_s$ and $\la^{1/2}/g_s$, respectively, while the 
higher order terms have negative powers of $\la$,
hence in the large $\la$ limit higher order corrections in \eqref{eq:W-expand}
are suppressed. Indeed we have checked that the inclusion
of higher order corrections does not change the plot significantly, and
the exact result \eqref{eq:genA}
is well approximated already at the next-to-leading order.
We have performed similar numerical checks
for various values of $N$ and $\la$ ($N,\la\gg1$)
and find a good agreement for all cases. 

\section{Conclusion}\label{sec:discuss}
We have computed the $1/\la$ expansion (or $\al'$-expansion of bulk type IIB string theory)
of Wilson loop VEV in the anti-symmetric representation
using the Sommerfeld expansion of Fermi distribution function. 
It would be very interesting to reproduce this result from the
computation of $\al'$-correction of the D5-brane action
in the $AdS_5\times S^5$ background.

There are many things to be studied further.
It is important to develop a  method to
compute both the $1/\la$ expansion and 
the $g_s$ expansion systematically.
In particular, it would be interesting to find the $1/\la$ expansion
of Wilson loop VEV in the symmetric representation
by the low temperature expansion of Bose distribution.
However, the integrand of the $\mu$-integral might have a singularity
corresponding to the onset of Bose-Einstein condensation.
It would be interesting to understand the analytic structure of the integrand
 in the case of symmetric representation (see \cite{Hartnoll:2006is} for a discussion).

Also, it is interesting to understand the convergence property of the expansion.
For the 1/2 BPS Wilson loop in the fundamental representation,
it is observed that the $\alpha'$-expansion
is {\it not} Borel summable \cite{Drukker:2000rr},
reflecting the fact that there are corrections of order $e^{-\rt{\la}}$
which is non-perturbative in $\al'$.  
It would be very interesting to understand the
Borel summability of $\alpha'$-corrections at fixed $g_s$
for Wilson loops in various representations.
On the other hand, the $g_s$ expansion of
1/2 BPS Wilson loops with fixed $\la$ seems to have a finite radius of convergence\footnote{For the Wilson loop in the fundamental representation,
we have checked the convergence of $g_s$ expansion
numerically using the result in \cite{Okuyama:2006ir}.},
which is consistent with the absence of Yang-Mills instanton corrections to Wilson loop
VEV in $\mathcal{N}=4$ SYM \cite{Pestun:2007rz}. 
We hope that the study of Wilson loops
in various representations will
provide us with precious information of the non-perturbative structure of string theory.

Finally, we would like to emphasize the importance of our findings.
It was reported in \cite{Faraggi:2014tna}
that there is a discrepancy
between the
computation on the field theory side and the
string theory side of the one-loop correction in $1/N$
\eqref{eq:log-corr} (see
\cite{Zarembo:2016bbk} for the 
 current status of this problem).\footnote{Actually, to understand the origin of this discrepancy was one of the motivation of this work.
}
However, before settling the issue of this problem of one-loop correction,
we have to compute the leading term at large $N$ as a function of 
$\la$, including all $1/\la$ corrections \eqref{eq:W-expand}.  
After having found the leading term \eqref{eq:W-expand},
one can try to study the one-loop correction in $1/N$, either
numerically or analytically, by subtracting the leading term
from the exact result \eqref{eq:genA}.
In this computation, 
it is important to subtract
the subleading terms in $1/\la$ since they
are rather large and cannot be neglected, as we have seen in section \ref{sec:compare}.
We believe that our result \eqref{eq:W-expand}
is an important first
step to resolve the issue of the one-loop discrepancy.
We leave the study of one-loop correction 
as an interesting future problem. 

\vskip8mm
\centerline{\bf Acknowledgments}
\vskip2mm
\noindent
The work of K.O. was supported in part by JSPS KAKENHI Grant Number
16K05316, and JSPS Japan-Hungary and Japan-Russia bilateral
joint research projects.
 

\end{document}